# Quantum Reconstructions as Stepping Stones toward *ψ-Doxastic* Interpretations?

Philipp Berghofer

University of Graz

philipp.berghofer@uni-graz.at

**Abstract**

In quantum foundations, there is growing interest in the program of reconstructing the quantum formalism from clear physical principles. These reconstructions are formulated in an operational framework, deriving the formalism from information-theoretic principles. It has been recognized that this project is in tension with standard *ψ-ontic* interpretations. This paper presupposes that the quantum reconstruction program (QRP) (i) is a worthwhile project and (ii) puts pressure on *ψ-ontic* interpretations. Where does this leave us? Prima facie, it seems that *ψ-epistemic* interpretations perfectly fit the spirit of information-based reconstructions. However, *ψ-epistemic* interpretations, understood as saying that the wave functions represents one's *knowledge* about a physical system, recently have been challenged on technical and conceptual grounds. More importantly, for some researchers working on reconstructions, the lesson of successful reconstructions is that the wave function does *not* represent objective facts about the world. Since knowledge is a factive concept, this speaks against epistemic interpretations. In this paper, I discuss whether *ψ-doxastic* interpretations constitute a reasonable alternative. My thesis is that if we want to engage QRP with *ψ-doxastic* interpretations, then we should aim at a reconstruction that is spelled out in non-factive experiential terms.

# 1. Introduction

The quantum reconstruction program (QRP) enjoys much popularity in the recently emerged field of quantum foundations (see, e.g., D'Ariano et al. 2017 and Chiribella & Spekkens 2016). The idea is that instead of taking the quantum formalism as a given and trying to make sense of it by reading off an ontology from the mathematics involved, we should aim at finding clear physical principles from which the formalism can be derived or reconstructed. The underlying rationale is that understanding quantum mechanics can best be achieved by identifying and analyzing the physical principles from which it can be derived. One motivation for QRP is that Einstein did something similar for special relativity. The mathematics underlying special relativity was discovered before Einstein, but only when Einstein succeeded in deriving the mathematics from meaningful principles – the light postulate and the principle of relativity – did special relativity emerge as a well-understood and broadly accepted scientific theory. It is to be noted that while QRP is currently highly popular among physicists working on quantum foundations, the program remains widely ignored among philosophers.[1] One reason why QRP might not be so popular among philosophers is that it seems to be in tension with so-called *ψ-ontic* interpretations (see Koberinski & Müller 2018). The most popular interpretations in philosophy – Bohmian mechanics, the many-worlds interpretation, and collapse theories – are all *ψ-ontic* interpretations. Furthermore, proponents of these interpretations like to refer to them as "quantum theories without observers" (Dürr & Lazarovici 2020, viii; Goldstein 1998), which does not square well will QRP since the latter operates in an *operational* framework. As discussed below, at present all successful reconstructions are reconstructions based on *information*-theoretic principles.

---

1  What I mean by this is that the quantum reconstruction program is often not even mentioned also in very recent influential works on philosophy of quantum mechanics such as (Adlam 2021, Dürr & Lazarovici 2020, Friebe et al. 2018, Knox & Wilson 2022, Maudlin 2019, Wallace 2021a). And when QRP is discussed in philosophical publications, the respective work is often (co-)authored by a physicist actively working on quantum reconstructions such as in the works of Bub (e.g., Bub 2004 and Bub & Pitowsky 2010), (Berghofer et al. 2021), or (Koberinski & Müller 2018). It is to be noted that when QRP is explicitly discussed in philosophy, it is typically well-received (Adlam 2022, Dickson 2015, Dunlap 2022, Felline 2016, Grinbaum 2006, 2007). The exceptions to this rule of being well-received are (Brown & Timpson 2006) and, with some qualifications, (French 2023).

Importantly, it is not the aim of this paper to motivate or defend QRP. Instead, I presuppose that QRP is a worthwhile project and that it indeed does put pressure on *ψ-ontic* interpretations. This raises the following question: If QRP is in tension with *ψ-ontic* interpretations, what kind of interpretation regarding the nature of the wave function does QRP support? The natural answer to this question is *ψ-epistemic* interpretations. And indeed, this is the alternative typically suggested in the literature (see, e.g., Koberinski & Müller 2018). However, there are several reasons to be suspicious about epistemic interpretations. First, *ψ-epistemic* interpretations, in the way they have been introduced in the influential work of (Harrigan & Spekkens 2010), have been prominently challenged by the PBR theorem (Pusey et al. 2012). Second, it has been noted that the view that the wave function exclusively represents one's knowledge about a physical system, but not the physical system itself, leads to conceptual problems since the concept of knowledge is factive (Luc 2023). Third, in this paper we are particularly interested in the idea that successful reconstructions suggest that the wave function does *not* represent objective facts about the world. Thus, we should not think about the wave function in terms of factive concepts such as knowledge.

Where does this leave us? In this paper, I suggest what I call *ψ-doxastic* interpretations as a way out. According to *ψ-doxastic* interpretations, the wave function neither represents an ontic state nor our knowledge/uncertainty about an underlying ontic state. Instead, it represents degrees of belief. The most prominent *ψ-doxastic* interpretation is QBism. The thesis of this paper is that if we want to engage QRP with *ψ-doxastic* interpretations, then we should seek to find a reconstruction of quantum theory that is spelled out in non-factive terms. More precisely, what I suggest in this paper is to entertain the possibility to reconstruct the quantum formalism from phenomenological principles spelled out in terms of *experience*. While this may seem outlandish, I will show that the principles suggested in Rovelli 1996, which is the work that started the quantum reconstruction program, can be reformulated by exploiting the phenomenological thesis that experience is necessarily *perspectival*. To be sure, my objective is not to offer a reconstruction (because I can't).

It is to motivate reconstructions in non-factive terms and to argue that reconstructions in terms of experience may be a viable option.

The paper is structured as follows. Section 2 clarifies how *ψ-doxasticism* can be motivated from the perspective of QRP. Section 3 introduces QBism, which is the most prominent *ψ-doxastic* interpretation. In Section 4, I discuss the possibility of reconstructing quantum theory from phenomenological principles about the perspectival character of experience.

**2. From QRP to *ψ-doxastic* interpretations**

The cornerstones of the quantum reconstruction program (QRP) have first been formulated by Carlo Rovelli (1996). Here, Rovelli argues that quantum mechanics needs to be based on a set of simple *physical* principles, suggesting concrete *information-theoretic* principles that could play such a role. Special relativity is identified as a role model for this program: a physical theory that has counter-intuitive consequences but is widely accepted since it conceptually rests on clear physical principles. This is how Rovelli captures the spirit, approach, and ambition of QRP:

> Quantum mechanics will cease to look puzzling only when we will be able to *derive* the formalism of the theory from a set of simple physical assertions ('postulates,' 'principles') about the world. Therefore, we should not try to *append* a reasonable interpretation to the quantum *mechanics formalism,* but rather to *derive* the formalism from a set of experimentally motivated postulates. (Rovelli 1996, 1639)

In the years to follow, the success of and booming interest in quantum information theory convinced more and more researchers that the notion of information is crucial for understanding the foundations of quantum mechanics. In the year 2000, Christopher Fuchs and Gilles Brassard co-organized a conference with the programmatic title "Quantum Foundations in the Light of Quantum Information." Fuchs' paper of the same name has been highly influential, summarizing the methodology of this project as reducing "quantum theory to two or three statements of crisp *physical* (rather than abstract, axiomatic) significance. In this regard, no tool appears to be better calibrated for a direct assault than quantum information theory" (Fuchs 2001). Soon after the conference, the first successful reconstruction was offered by (Hardy 2001). Several others followed, e.g., Chiribella et al. (2011), Dakic & Brukner (2011), Masanes et al. (2013), Goyal (2014), Höhn (2017), and Höhn & Wever (2017), and the reconstruction program continues to shape the field of quantum foundations (as exemplified by Chiribella & Spekkens 2016 and D'Ariano et al. 2017).[2] As mentioned above, it is certainly a virtue of this program that it is based on a simple and convincing idea.

This idea is that instead of taking the quantum formalism as a given and trying to make sense of it by contemplating the ontological status of the mathematical terms involved, such as, most prominently, the wave function, we should be looking for foundational physical principles from which the formalism can be derived or reconstructed.[3] The point is that the formalism of quantum mechanics is couched in highly technical terms that complicate a direct interpretation. The spirit of the reconstruction program is that instead of asking how to ontologically interpret the mathematics, we want to know where the mathematics comes from. Why is nature so successfully described by the mathematics of complex Hilbert spaces? The idea is that this question can only be answered by deriving or reconstructing the formalism from principles that have a clear physical meaning.

---

[2] Of course, this is only a very rough glimpse at the history of quantum reconstruction. In fact, operational axiomatizations can be traced back to von Neumann himself and his joint work with Garrett Birkhoff on quantum logic. It has been pointed out in this context that von Neumann confessed to Birkhoff that he did not "believe in Hilbert space anymore" (D'Ariano et al. 2017, 2; Grinbaum 2017). The quantum reconstruction program also significantly drew on developments in probability theory. The QBist reconstructions were significantly influenced by the work of Bruno de Finetti (see, e.g., Fuchs 2001). The reconstructive work offered in Goyal et al. 2010 was influenced by Cox' derivation of probability theory from Boolean algebra. Furthermore, although Rovelli 1996 can be called the proper beginning of QRP, similar ideas have been developed independently in Popescu & Rohrlich 1994 and particularly in Zeilinger 1999.

[3] For instance, Fuchs and Stacey argue that in contemporary philosophy of quantum mechanics "the strategy has been to reify or objectify all the mathematical symbols of the theory and then explore whatever comes of the move" (Fuchs & Stacey 2019, 136). QRP, by contrast, is considered a project of "taking a more physical and less mathematical approach" (Masanes et al., 2013, 16373). At this point, we need to emphasize the substantial distinction between wave function realism and *ψ-ontic* interpretations. Wave function realism is the view that the wave function is physically real. An interpretation is *ψ-ontic* if it says that the wave function represents an ontically real state. Wave function realism is thus a much stronger claim, particularly since wave functions are mathematically defined as highly abstract vectors in Hilbert space. Wave function realism and *ψ-ontic* interpretations align well, but while the former implies the latter, this is not true vice versa. Thus, when I say that the most popular interpretations in philosophy – Bohmian mechanics, the many-worlds interpretation, and collapse theories – are all *ψ-ontic* interpretations, I do not imply that they are also committed to wave function realism.

> In short, the postulates of quantum theory impose mathematical structures without providing any simple reason for this choice: the mathematics of Hilbert spaces is adopted as a magic blackbox that 'works well' at producing experimental predictions. However, in a satisfactory axiomatization of a physical theory the mathematical structures should emerge as a consequence of postulates that have a direct physical interpretation. By this we mean postulates referring, e.g., to primitive notions like physical system, measurement, or process, rather than notions like, e.g., Hilbert space, $C^*$-algebra, unit vector, or self-adjoint operator. (D'Ariano et al. 2017, 1)

Compare this sentiment with the approach most popular in philosophy of physics: "A precisely defined physical theory [...] would never use terms like 'observation,' 'measurement,' 'system,' or 'apparatus' in its fundamental postulates. It would instead say precisely *what exists and how it behaves*" (Maudlin 2019, 5). The interpretations of quantum mechanics most popular in philosophy, i.e., Bohmian mechanics, the many-worlds interpretation, and objective collapse theories, are so-called "quantum theories without observers" (Dürr & Lazarovici 2020, viii; Goldstein 1998). The dominant view here is that the way quantum mechanics is taught and understood in physics textbooks is not only misleading but plainly unscientific. This is precisely because in textbook quantum mechanics, "measurement" is a central notion. The dominant strand in philosophy of quantum mechanics seeks to "develop an objective description of nature in which 'measurements' are subject to the same laws of nature as all other physical processes," insisting that "the goal of physics must be to formulate theories that are so clear and precise that any form of interpretation […] is superfluous" (Dürr & Lazarovici 2020, viii). Accordingly, there is some tension between the operational framework in which QRP operates and mainstream analytic philosophy of physics.

Furthermore, the mainstream approaches in philosophy are *ψ-ontic* interpretations. This terminology has been introduced by Harrigan and Spekkens (2010), contrasting *ψ-ontic* with *ψ-epistemic* interpretations. In this paper, we use this terminology in the following informal sense: an interpretation is *ψ-ontic* if it regards the wave function as representing an ontic state, while it is *ψ-epistemic* if it regards the wave function representing our knowledge/uncertainty about the state.[4] QRP is in tension with *ψ-ontic* interpretations in the following sense: If successful reconstructions succeed in capturing the spirit of quantum mechanics and the nature of the wave function in information-theoretic terms, why should we also believe that the wave function has the additional role of representing physical reality? From the perspective of QRP, ontological interpretations of the quantum state seem superfluous. In this context, it is to be noted that proponents of QRP "characterize quantum theory *as a theory of information*" (D'Ariano et al. 2017, 5), and then it is natural to assume that the lesson to be learned from QRP is to interpret "quantum theory as a theory about the representation and manipulation of information, which then becomes the appropriate aim of physics, rather than a theory about the ways in which nonclassical waves or particles move" (Bub 2004, 243).[5] For instance, Höhn and Wever understand the implications of their reconstruction as follows:

---

4  Harrigan and Spekkens introduce this terminology within their ontological model framework, which is tailor-made for hidden-variable theories. Although this ontological framework is consistent with QRP, none of the successful reconstructions mentioned above is formulated within this framework. An important feature of this framework is that any *ψ-complete* model must be *ψ-ontic*. In other words, if a model is *ψ-epistemic*, it must be *ψ-incomplete*. Importantly, when proponents of QRP stress that successful reconstructions speak in favor of *ψ-epistemic* interpretations, they typically do *not* want to say that the wave function is incomplete. In such contexts, the terminological distinction between ontic and epistemic interpretations is always understood in the informal sense as introduced above.

5  For more details on Bub's information-theoretic interpretation of quantum mechanics, see (Bub 2018), (Dunlap 2022), and (Janas et al. 2022). It seems that particularly the views expressed in the second edition of Bub's *Bananworld* cohere nicely with my approach. For some of the subtle differences between the first and second edition of *Bananaworld*, see (Dunlap 2022).

> [T]he successful reconstruction from this perspective underscores the sufficiency of taking a purely operational perspective, addressing only what an observer can say about the observed systems, in order to understand and derive the formalism of quantum theory. Ontic statements about a reality underlying the observer's interactions with the physical systems are unnecessary. (Höhn & Wever 2017).

The objective of this paper is not to defend QRP or the claim that QRP puts pressure on $\psi$-*ontic* interpretations. I will do so in a companion paper. The objective is to investigate what it would mean for the proponents of QRP if it were true that they should look for alternatives to $\psi$-*ontic* interpretations. The obvious alternative is $\psi$-*epistemic* interpretations. After all, $\psi$-*ontic* interpretations have been introduced in distinction to $\psi$-*epistemic* interpretations. Furthermore, reconstructions are typically formulated in terms of information or knowledge. And indeed, this option of going $\psi$-*epistemic* has been suggested by (Koberinski & Müller 2018).

One thing to be noted about $\psi$-*epistemic* interpretations is that even strong proponents of $\psi$-*ontic* interpretations readily admit that epistemic interpretations have at least one crucial virtue. This is that epistemic interpretations avoid the measurement problem. "Any approach according to which the wave function is not something real, but represents a subjective information, explains the collapse at quantum measurement perfectly: it is just a process of updating the information the observer has" (Vaidman 2014, 17). According to textbook quantum mechanics, the wave function is governed by the deterministic Schrödinger equation *unless a measurement takes place*. Upon measurement, the wave function "collapses" and the Schrödinger dynamics is replaced by a genuinely stochastic process. This is particularly puzzling if you consider the collapse a physical process. Why should nature care about whether a measurement is performed? $\psi$-*epistemic* interpretations easily avoid all these problems. In this picture, the wave function does not represent an ontic state but the observer's knowledge about the state, and the collapse is not a physical process but an update in knowledge. The wave function collapses in the sense that the uncertainty is gone: once the measurement has been performed, the observer knows, for instance, the position of the electron.

Now, the problem is that *ψ-epistemic* interpretations have been challenged on technical and conceptual grounds. For instance, the PBR theorem shows that allowing for wave function overlap on state space contradicts quantum theory, which means that *ψ-epistemic* interpretations as introduced by (Harrigan & Spekkens 2010) are untenable. Furthermore, it has been noted that the view that wave function exclusively represents one's knowledge about a physical system, but not the physical system itself, leads to conceptual problems (Luc 2023). This is because knowledge is a *factive* concept: If one knows that *p*, then *p* is true. Thus, if we say that the wave function represents one's knowledge about a given system, we make a representational claim about the system itself. Does this imply that looking for alternatives to *ψ-ontic* interpretations leads to a dead end? This is not the case. It is not the case because, as noted by several authors, *ψ-ontic* and *ψ-epistemic* interpretations are not the only options. An alternative to *ψ-ontic* and *ψ-epistemic* interpretations are what I call *ψ-doxastic* interpretations.[6] A *ψ-doxastic* interpretation says that the wave function neither represents an ontic state nor the knowledge about the underlying ontic state but the agent's belief or judgment. Of course, this is a very uncommon way of interpreting a physical theory. Typically, we would say that the basic objects of a physical theory represent physical objects – as is the case in classical mechanics. However, the basic objects of quantum mechanics are not point particles but wave functions. While the point particles of classical mechanics live in three-dimensional Euclidean space, wave functions in quantum mechanics are vectors defined in highly abstract mathematical Hilbert space, and it is well-known how notoriously difficult it is to make sense of them. One consistent way to do so is by interpreting them doxastically. This is the project of QBism.

---

6    The term is anticipated in DeBrota & Stacey (2019) and Luc (2023). The only published work I know of in which it is explicitly used is Ruebeck et al. (2020).

## 3. QBism & QRP

### 3.1. QBism: The basics

The distinctive idea of QBism is to apply a personalist Bayesian account of probability, as developed by Bruno de Finetti, to quantum probabilities (Fuchs et al. 2014). This means that probabilities in quantum mechanics are interpreted not as objective but as subjective probabilities. Quantum states do "not represent an element of physical reality but an agent's personal probability assignments, reflecting his subjective degrees of belief about the future content of his experience" (Fuchs & Schack 2015, 1). This is to say that the wave function neither is a physical object nor does it represent a physically real state but it is a tool that encodes the experiencing subject's expectations about her future experiences. QBism has a normative dimension in the sense that the Born rule is viewed as a *normative* constraint that "functions as a consistency criterion which puts constraints on the agent's decision-theoretic beliefs" (Schack 2023, 146).[7] The focus on experience also manifests when it comes to the concept of measurement. For the QBist, a measurement is an act of the subject on the world and the outcome of a measurement is the very experience that results from this process (see DeBrota & Stacey 2019). Following Wheeler, QBists reject standard scientific realism[8] and insist that "reality is *more* than any third-person perspective can capture" (Fuchs 2017, 113). Focusing on the relationship between the experiencing subject and the experienced world, they propose a kind of realism that has been labeled a "participatory realism" (Fuchs 2017).[9]

---

[7] However, it must be emphasized that QBists explicitly deny that the wave function can represent what the subject *should* believe. Wave functions and quantum probabilities are purely subjective. I believe that interpreting quantum probabilities as objective degrees of epistemic justification would be an interesting option that preserves the virtues of QBism but avoids some of its problems. Such an interpretation would also cohere nicely with the approach suggested in Section 4.

[8] An anonymous reviewer of this journal pointed out that the scientific realism debate is typically formulated in terms of one's attitude toward so-called unobservable entities such as atoms, electrons, and fields: scientific realists claim that we are justified in believing in the existence of such entities postulated by our best scientific theories; anti-realists deny this. It is crucial to emphasize that QBists are not anti-realist in this sense. QBists have absolutely no hesitation to accept that entities such as atoms, quarks, or electromagnetic fields exist (see Pienaar 2023). Of course, they are also not anti-realist in the sense that they deny the existence of an external world. To the contrary, the existence of an external world is presupposed by QBism (Fuchs 2023, 92). What sets QBism apart from standard realist interpretations of quantum mechanics is that it denies that the wave function represents reality. Accordingly, the label "anti-representationalist" fits QBism better than "anti-realist." In this light, what QBists (should) criticize about a hidden-variable theory such as Bohmian mechanics is not that it introduces quantities that are unobservable (to us), but that the structure of the theory violates the following principle: "a theory shouldn't make distinctions that it cannot empirically honor" (Carrier 2012, 28; my translation). This has been the driving idea behind Einstein's development of relativity theory (Carrier 2012) as well as Heisenberg's quantum mechanics.

QBism and QRP are intimately related both historically and systematically. Regarding the development of both projects, Fuchs is not only a co-founder and the main proponent of QBism, as we have seen he is also a founding figure of the quantum reconstruction program (Fuchs 2001). In general, QBists are actively involved in QRP (Schack 2003, Appleby et al. 2017, DeBrota et al. 2020). Systematically, QBists believe that their approach to the nature of the wave function (quantum state) is well-motivated by the achievements of quantum information theory.

> A quantum state encodes a user's beliefs about the experience they will have as a result of taking an action on an external part of the world. Among several reasons that such a position is defensible is the fact that any quantum state, pure or mixed, is equivalent to a probability distribution over the outcomes of an informationally complete measurement. Accordingly, QBists say that a quantum state is conceptually no more than a probability distribution. (DeBrota & Stacey 2019).

This exemplifies the close connection between reconstruction and interpretation and how the success of QRP can be understood as motivating and supporting approaches that are in tension with $\psi$-ontic interpretations. However, recently Fuchs and Stacey have expressed discomfort with the development of QRP (Fuchs & Stacey 2016). I return to why there might be a tension between QBism and *informationally* reconstructing quantum mechanics in Subsection 3.4.

---

This is how Heisenberg opens his 1925 article, marking the beginning of modern quantum mechanics: "The objective of this work is to lay the foundations for a theory of quantum mechanics based exclusively on relations between quantities that are in principle observable" (Heisenberg, as cited in Rovelli 2021, 20).

9   For more details on QBism and how it can be related to various philosophical disciplines, see Timpson 2008, 2013; Glick 2021; and Berghofer 2023 and Berghofer & Wiltsche (forthcoming).

## 3.2. Common misconceptions

QBism is the currently best-developed interpretation of quantum mechanics that embraces the Copenhagen spirit that "*experience* is fundamental to an understanding of science" in the sense that "quantum mechanics is a tool anyone can use to evaluate, on the basis of one's past experience, one's probabilistic expectations for one's subsequent experience" (Fuchs et al. 2014, 749).[10] Regarding the wave function, QBists are clear that the wave function should not be reified or objectified and they reject the idea that the wave function represents objective reality. This is why it is often believed that QBism is a *ψ-epistemic* interpretation (see, e.g., Koberinski & Müller 2018). Importantly, this is misleading.[11] As noted above, this terminological distinction has been introduced by Harrigan & Spekkens (2010). They call an interpretation "*ψ-ontic* if every complete physical state or *ontic state* in the theory is consistent with only one pure quantum state; we call it *ψ-epistemic* if there exist ontic states that are consistent with more than one pure quantum state" (126). They point out that this means that "[o]nly in the latter case can the quantum state be considered to be truly epistemic, that is, a representation of an observer's knowledge of reality rather than reality itself" (126).

But QBism is more radical than that. In QBism quantum states neither represent an underlying ontic state, nor our knowledge/uncertainty of an underlying ontic state. Instead, they are interpreted as representing the subject's beliefs about her future experiences (DeBrota & Stacey 2019, 10). QBists rightly emphasize that this is a *doxastic* and not an *epistemic* interpretation of the quantum state/wave function. This is because knowledge is a factive notion. If one knows that $p$, then $p$ is the case. Importantly, this is *precisely* why QBism avoids the PBR theorem. While the PBR theorem challenges *ψ-epistemic* interpretations, it is silent on the QBist claim that wave functions represent degrees of beliefs about one's future experiences (DeBrota & Stacey 2019, Glick 2021, Hance et al. 2022).

---

10  "In many ways, quantum Bayesianism represents the *acme* of certain traditional ways of thinking about quantum mechanics (broadly speaking, Copenhagen-inspired ways). If one hopes to defuse the conceptual troubles over collapse and nonlocality by conceiving of the quantum state in terms of some cognitive state, then the only satisfactory way to do so is by adopting the quantum Bayesian line." (Timpson 2013, 7f.)

11  It is to be noted, however, that in its early formulation Quantum Bayesianism has been introduced as an epistemic interpretation, explicitly arguing that "quantum states are states of knowledge" (Caves et al. 2002). More than a decade later, they clarified why QBism should abandon this view (Fuchs et al. 2014, 753). As pointed out by Stacey, this can be understood as a "shift of interpretation" from an objective-Bayesian interpretation to a personalist-Bayesian one (Stacey 2019, 6).

Unfortunately, in the literature the PBR theorem is often misunderstood as ruling out any interpretation that is not *ψ-ontic* (e.g. Maudlin 2019, 83-89). This means overlooking how the QBist escapes the PBR theorem. In fact, the implications of PBR should be understood as revealing that QBism is one of the main alternatives to *ψ-ontic* interpretations.

One might wonder whether there is a tension between two of the central claims of QBism. On the one hand, it is emphasized that the wave function does *not* represent reality and that quantum mechanics is *not* a descriptive theory that tells us how external reality evolves in time. On the other hand, the quantum formalism is considered the most effective tool we have at our disposal to correctly predict what we will experience next. This leads us to the following question raised by an anonymous reviewer of this journal: How can there be an outside world (as QBism acknowledges), in which the wave function does not represent features of the world, and still the wave function is successful in predicting and guiding an agent in the world? This, precisely, is the million-dollar question. QBists not only recognize that they owe us an answer to this question, they insist that the very goal of their research project is to "reverse engineer" from the quantum formalism to the specifics of reality. Here is how Fuchs puts it:

> "Indeed as emphasized in the Introduction, from its earliest days the very goal of QBist research has been to distill a statement about the character of the world from the fact that the gambling agents within it should use the quantum formalism. Why would it be so? Whatever the answer turns out to be, it will be a statement about the particulars of reality. If the world were different in character, then the agents within that world would be better advised to use something other than quantum theory for their gambles. […] What QBism aims for is to reverse engineer from the formalism to a characterization of an ontology, while never straying from the progress it has made by viewing quantum theory as an addition to decision theory. This reverse engineering remains an active research program—a sign that QBism is a living subject." (Fuchs 2023, 97f.)

The idea is that if the quantum formalism is a tool that can be successfully applied to reality, then from the specifics of the tool we should be able to infer substantial claims about the specifics of reality. As QBists readily admit, their attempt to provide an answer to this question remains work in progress. One lesson, for them, is that measurements do not reveal pre-existing or pre-determined properties. From this they infer that we do not live in a block universe (Fuchs 2015; 2023). This is also why QBists insist that they are not instrumentalists. Their goal is to say something substantial about reality. They *do* believe that quantum mechanics teaches us important lessons about the structure of reality. However, in their view, this can only happen in a very indirect way. The main reason why QBists insist that the wave function cannot be interpreted as representing reality, and instead must be interpreted as representing only degrees of belief, is a combination of two factors. First, they believe that several results of quantum information theory, such as the quantum no-cloning theorem, push them in this direction (Fuchs 2023, 104). Second, they believe that any alternative to what I call a *ψ-doxastic* interpretation[12] inevitably forces one to accept non-locality (Fuchs et al. 2014). Of course, both claims can be and have been contested. However, it is not the objective of this paper to defend QBism. Instead, two of my objectives are to clarify the relationship between QBism and QRP (Sections 3.3 & 3.4) and to suggest a novel way of reconstructing that might better fit QBism (Section 4.2).

    One important step toward a better understanding of the relationship between QBism and QRP taken in this paper was to distinguish between *ψ-epistemic* and *ψ-doxastic* interpretations. One question I will address at the end of the paper is whether there is anything "in between" epistemic and doxastic interpretations. Thus, although my aim of this paper is not to criticize or defend QBism, I will critically examine a fundamental but virtually never discussed implicit assumption of QBism, namely that if the wave function neither represents an ontic state, nor our knowledge of an underlying ontic state, then the wave function must represent degrees of belief. In Fuchs 2002,

---

12  Fuchs says that "quantum states should be understood epistemically" (2023, 104), but we have already clarified that it is useful to make a terminological distinction between *ψ-epistemic* and *ψ-doxastic* interpretations.

Fuchs captures the development of his thinking about quantum states as follows: "*Knowledge → Information → Belief → Pragmatic Commitment*." For an agent- or experience-centered approach such as QBism, the notions of knowledge and information are problematic because they are factive concepts. However, the notions of belief and pragmatic commitment, for many researchers and critics of QBism, are simply too subjective. How can objectivity enter science if our most fundamental scientific theory represents the subject's degrees of belief? I think the alternative missing in Fuchs' list "*Knowledge → Information → Belief → Pragmatic Commitment*" is "epistemic justification" in between "information" and "belief." I will discuss this alternative in Section 5.

### 3.3. How QBism can benefit from reconstructions

Even proponents of standard realist interpretations typically agree that QBism delivers a consistent interpretation of quantum mechanics that avoids problems surrounding the apparent collapse of the wave function and non-locality (Vaidman 2014, 17f.). However, QBism faces a number of philosophical challenges. Most of these challenges concern the instrumentalist flavor of QBism. Although QBists explicitly insist that QBism must be understood "as being part of a realist program, i.e., as an attempt to say something about what the world is like, how it is put together, and what's the stuff of it" (Fuchs, 2017, 117), QBism has an instrumentalist touch since it denies that quantum mechanics is a theory that represents physical states and describes how these states evolve. One particularly prominent charge in this context is that QBism cannot *explain* quantum phenomena (or the behavior of the world in general). More precisely, the worry is that QBism "would rob quantum theory of explanatory power which it nonetheless seems to possess" (Timpson 2008, 581). Let's call this the explanatory deficiency challenge. This worry is legitimate. For instance, we would like to know why there is an interference pattern when we fire electrons toward a double slit. Bohmians, for instance, seem to have a reasonable explanation. The basic ontology of quantum mechanics is particles, these particles evolve according to certain equations of motion, it follows from these equations that we will observe an interference pattern. This is the kind of dynamical explanation we are used to from classical mechanics, and it works. When we ask the QBists to explain the double slit experiment, we don't get a similarly physical explanation (DeBrota & Stacey 2019, Section 7). Now, in light of the above, the following proposal arises naturally: QBism should aim at explaining the quantum phenomena by means of a suitable reconstruction. To my knowledge, QBists have explicitly addressed how QRP might help QBists to answer the explanatory deficiency challenge only once in passing in (DeBrota & Stacey 2019, Section 19). This is a bit surprising because, as mentioned above, QBists are heavily involved in QRP. However, raising and answering a worry such as the explanatory deficiency challenge is a genuinely philosophical undertaking, so here we are.

It is useful, once again, to consider an analogy to special relativity. What is the physical explanation for length contraction (aka Lorentz contraction)? Why is it that a body's length is measured to be shorter when it moves? If we follow Einstein, the explanation is that this is a consequence of his two postulates. If we follow Minkowski, the explanation is that it is a consequence of the geometry of spacetime. We accept this kind of explanation because we accept and understand the underlying principles. However, this kind of explanation is different from the dynamical one we know from classical mechanics. And not everybody was convinced by it. Most notably, Lorentz himself did not consider it a sufficiently physical explanation. Here is how Rovelli summarizes this attitude:

> The physical interpretation proposed by Lorentz himself (and defended by Lorentz long after 1905) was a physical contraction of moving bodies, caused by complex and unknown electromagnetic interaction between the atoms of the bodies and the ether. It was a quite unattractive interpretation (and remarkably similar to certain interpretations of wave function collapse as presently investigated!). (Rovelli 1996, 1639f.)

Indeed, currently popular "modificatory" interpretations are somewhat reminiscent of Lorentz' approach. For instance, Bohmian mechanics modifies the formalism of quantum mechanics, introduces definite and pre-determined particle trajectories as "hidden variables," and has the unobservable wave function to somehow act upon the positions of the particles without being acted upon by the particles.[13] The promise of the quantum reconstruction program, in this light, is to offer an approach such that quantum phenomena can be explained by being traced back to physically meaningful principles without having to modify the formalism and falling back on stipulating the existence of unobservable entities.

This is also the right place to very briefly address the charge of instrumentalism more generally. Broadly speaking, instrumentalism is the view that science is only in the business of prediction, not explanation. Scientific theories, in this view, are black boxes whose outputs are predictions and whose value is exclusively determined by the accuracy of these predictions. It is obvious that QRP is not an instrumentalist project. Everyone in the reconstruction community acknowledges the predictive power of quantum mechanics. The ambition is not to reconstruct the formalism in order to make better predictions, but to gain a better understanding of the theory. In this sense, QRP is all about explanation.[14] Accordingly, QBism paired with QRP cannot be criticized for being merely instrumentalist. The question is not whether QBism is in the business of explanation. The question is whether we find the explanations it is capable of offering satisfactory. The persuasiveness of these explanations crucially depends on the respective reconstruction. This is to say that the best way for QBists to prove the explanatory power of QBism is to find a convincing and suitable reconstruction.

---

13  It does not come as a surprise, then, that Lorentz was one of the few physicists who "showed much sympathy for de Broglie's attempt to develop a theory of particle trajectories" (Dürr & Lazarovic 2020, 76) at the famous 1927 Solvay conference. However, as an anonymous reviewer of this journal pointed out: "There is a sophisticated argumentation as to why, and in what sense, the wave function in Bohmian mechanics guides the particles in a nomological sense" (see also Goldstein 2021, Section 17). This serves as a reminder that Bohmians are not forced to accept wave function realism. However, they are forced to accept that physically distinct states cannot be distinguished by observation. As Kofler and Zeilinger put it: "While the testable predictions of Bohmian mechanics are isomorphic to standard Copenhagen quantum mechanics, its underlying hidden variables have to be, in principle, unobservable. If one could observe them, one would be able to take advantage of that and signal faster than light, which – according to the special theory of relativity – leads to physical temporal paradoxes" (Kofler & Zeilinger 2010, 474).
14  See in this context also Dickson's notion of "sensible instrumentalism" (Dickson 2015).

**3.4. Tensions between QBism and QRP**

Above we noted that in its early days Quantum Bayesianism was introduced as an epistemic interpretation according to which "quantum states are states of knowledge" (Caves et al. 2002). By now, QBists have renounced this view. Why? Because QBists shifted from an objective-Bayesian understanding of quantum probabilities to a personalist one. Quantum states do not represent the subject's knowledge but the subject's degrees of belief. Why is this important? Knowledge is a factive notion but belief is not. If the subject knows that $p$, $p$ is the case. Accordingly, if we say that the quantum state represents the subject's knowledge about the world, this suggests that the quantum state is about an underlying ontic state the subject knows about. But according to QBism, the quantum state is not about the world but about what the subject expects to experience next. Of course, this is not to say that QBism denies that there is an observer-independent world. It simply rejects the notion that quantum mechanics should be understood as representing an observer-independent world. While QBists have explicitly renounced the knowledge view and terminology (Fuchs et al. 2014, 753), they still make heavy use of the "information" terminology in their reconstructions (Appleby et al. 2017, DeBrota et al. 2020). Information, however, in the everyday use of the term, is also a factive notion and closely linked to knowledge (see Timpson 2013, 12).[15]

    This means that there is some tension between QBism and how QRP typically proceeds. This is because reconstructive projects focus on and are formulated in terms of factive concepts such as "knowledge" or "information." As Höhn & Wever put it: "In particular, we take the quantum state

---

15  It is to be noted, however, that QBists seem to be aware that information is not the ideal term for their enterprise (Fuchs in Crease & Sares 2021, 544; Fuchs 2017, 120), so this might be a terminological convenience.

to represent the observer's 'catalog of knowledge' about the observed system(s), rather than an intrinsic state of the latter" (Höhn & Wever 2017, 1). However, it seems that from a QBist perspective it would be advantageous to spell this out in terms of non-factive mental states such as belief and experience. Interestingly, QBists have never voiced the ambition to do so. Although they have expressed their discomfort with the development of QRP (Fuchs & Stacey 2017), this criticism has concerned the lack of a purely physical non-information-theoretic principle *in addition* to information-theoretic principles. To my knowledge, they have never opted for formulations in purely non-factive terms and their own reconstructions do not stress this point either. However, it has been noted that there are many systematic similarities between philosophical approaches to the nature of experience (particularly in the phenomenological tradition) and the cornerstones of information-based reconstructions (see Berghofer et al. 2021). Furthermore, it has been noticed that while classical mechanics is typically considered to fit nicely with our physical intuitions, it does not fit at all with how we experience the world (Goyal 2023). However, when it comes to quantum reconstructions, it seems that in surprisingly many cases we can more or less straightforwardly reformulate certain foundational principles in experiential terms. In the following section we first introduce phenomenological teachings about the perspectival character of experience (Section 4.1) and then give an outline of how foundational information-theoretic principles could be reformulated in experiential terms (4.2).

## 4. Toward a phenomenological reconstruction of quantum theory

### 4.1. The perspectival character of experience

Phenomenology is the study of appearances, i.e., the study of experience and of objects *as objects of experience*. In the Husserlian tradition, this endeavor is understood as a *descriptive* and *eidetic* study of consciousness. This is to say that consciousness is studied from the first-person perspective and that the objective of this study is *not* to report one's current mental life and to find contingent empirical truths but to unveil a priori laws about the structure of consciousness. One of Husserl's main contributions to a proper phenomenological analysis of experience is the disclosure of what he calls the *horizontal structure of experience*. In this context, Husserl shows that perceptual experiences are genuinely *perspectival*. As we will see, this means that each and every experience only provides a limited perspective on the experienced object. These perspectives are *complementary* in the sense that different perspectives shed new light on the experienced object but cannot be acquired simultaneously. Importantly, in the phenomenological tradition this is not understood as a shortcoming of contingent human sensory limitations, but as a necessary feature of any subject experiencing a transcendent world. Other possible beings may experience the world in richer but still necessarily limited and complementary perspectives. A purely objective view from nowhere is considered an idealization that is in principle impossible.

Let's proceed step by step. The first thing to note is that according to phenomenological analyses perceptual experiences always and necessarily go beyond what is sensuously given.[16] This can be illustrated as follows: Assume that you are looking at a cup of coffee. At first glance, what presents itself to you in experience is a three-dimensional object in space. However, a closer examination reveals that what is really sensuously given to you is not simply a cup and its content, but only *one single profile of the object*, its current front side. Of course, you could wander around and make the current back side the new front side, and vice versa. But this doesn't change the fact that the cup is always given *in perspectives* and that, more generally, the objects of perceptual experiences always and necessarily have more parts, functions, and properties than can be actualized in one single intentional act.

---

16  Here, many of Husserl's insights are in agreement with the findings of early experimental psychologists such as Gestalt psychologists and the members of the Graz school. These ideas have been picked up in the recent movement of experimental phenomenology (see Albertazzi 2013). Although neglected for a long time, in the analytic tradition, there have recently been attempts to capture this distinctive character of perceptual experiences (e.g., Church 2013, 50). Particularly notable works in this context that blur the artificial distinction between analytic philosophy and phenomenology are (Madary 2017) and Smith (2010).

Thus, a closer look at how physical objects appear to us reveals that our intentions[17] toward them always "transcend" or "go beyond" what is directly sensuously given to us. There is a describable difference between what is meant through a particular perceptual act (e.g., that there is a cup of coffee in front of you) and what is sensuously given (the object's facing side with its momentarily visible features). This discrepancy, for the phenomenologist, is *not* a problem in need of resolution. Instead, the fact that our perceptual intentions always transcend the sphere of direct givenness is to be treated as a phenomenologically discoverable feature of experience itself. The perspectival character and horizonal structure of experience are not mere flaws in human perception but fundamental characteristics of experience.[18]

When Husserl discusses the perspectival character of perception, he emphasizes not only that perception is incomplete but also how physical objects always manifest from a particular viewpoint.

> All orientation is thereby related to a null-point of orientation, or a null-thing, a function which my own body has, the body of the perceiver. And again, the perspectival mode of givenness of every perceptual thing and of each of its perceptual determinations – on the other hand, also of the entire unitary field of perception, that of the total spatial perception – is something new. The differences of perspective clearly are inseparably connected with the subjective differences of orientation and of the modes of givenness in sides.[19] (Husserl 1977, 121)

A further aspect of perception is that previous experiences shape the way we perceive. Perception is not a faculty that allows us to see the world independent from our history, background beliefs, etc.

---

17  Here we use "intention" in the phenomenological sense according to which intentionality is an essential feature of consciousness, denoting the "aboutness" or "directedness" of our mental acts. In this sense, intentions can be understood as mental representations.

18  "Necessarily there always remains a horizon of determinable indeterminateness, no matter how far we go in our experience, no matter how extensive the continua of actual perceptions of the same thing may be through which we have passed." (Husserl 1982, 95).

19  It is interesting to see that the phenomenologically minded mathematician and physicist Hermann Weyl, father of the gauge principle which is one of the cornerstones of modern physics, basically makes the same claim, intentionally using phenomenological terminology (quoted and discussed in Ryckman 2005, 131).

To put it differently, "experience is not an opening through which a world, existing prior to all experience, shines into a room of consciousness; it is not a mere taking of something alien to consciousness into consciousness" (Husserl 1969, 232). This aspect of perception is closely related to discussions about the theory-ladenness of perception.

Although for Husserl experiences play an epistemologically foundational role, being a source of immediate justification as well as constituting our ultimate evidence, he is well aware that experiences are not windows to the world through which we see how the world is in itself thoroughly objectively. Instead, experiences present their objects in a certain way that at least partly depends on subjective factors such as previous experiences, background beliefs, etc. To put it differently, the objects we experience and think about do not have an objective sense that is for us to be discovered. Instead, we ourselves constitute the sense of the objects we engage with. This is to say that a view from nowhere at the world is in principle impossible. Regarding the relationship between sense and constitution, Moran provides the following summary:

> For Husserl, sense is not simply something outside us that we apprehend, it is something that is 'constituted' or put together by us due to our particular attitudes, presuppositions, background beliefs, values, historical horizons and so on. In short, phenomenology is a reflection on the manner in which things come to gain the kind of *sense* they have for us. (Moran 2012, 52)

Accordingly, we cannot achieve an objective view on the world, our experiences are necessarily incomplete and perspectival, by engaging with the world we constitute and thereby change the sense of the objects we encounter, and we only have limited knowledge of the present and the future. It is a commonplace in phenomenology that a purely objective third-person perspective is unreachable (see, e.g., Berghofer 2020 and Khalili 2022). As Zahavi puts it: "There is no pure third-person perspective, just as there is no view from nowhere. This is, of course, not to say that there is no third-person perspective, but merely that such a perspective is, precisely, a perspective from somewhere. It is a view that *we* can adopt on the world" (Zahavi 2019, 54). This resonates well with the QBist mantra that "reality is *more* than any third-person perspective can capture" (Fuchs 2017, 113).[20]

To sum up, experiences are perspectival in the sense that experiences provide limited perspectives. There always are infinitely many different perspectives possible that are complementary in the sense that they shed new light on the experienced object but cannot be acquired simultaneously. The process of experiencing can be understood as a process of anticipation and fulfillment (Madary 2017) in the sense that our experiences always have a horizon of unfulfilled intentions. When I look at the cup in front of me, I have certain anticipations of how the object looks from different angles. I can gain these new perspectives but when I do, I lose the former. In what follows, I discuss the possibility that certain information-theoretic principles that play a prominent role in current reconstructions could be reformulated in experiential terms. As discussed above, such reformulations in non-factive terms are desirable if you prefer *ψ-doxastic* over *ψ-epistmic/ontic* interpretations (as is the case for QBism).[21]

**4.2. Toward phenomenological reconstructions**

As noted above, Rovelli's highly influential paper "Relational Quantum Mechanics" (1996) is the work that marks the beginning of the quantum reconstruction program. Here Rovelli identifies two information-theoretic principles that he considers to capture the essence of quantum mechanics. Here they are (Rovelli 1996, 1657f.):

> "*Postulate 1* (Limited information). There is a maximum amount of *relevant information* that can be extracted from a system."

> "*Postulate 2* (Unlimited questions). It is always possible to acquire *new* information about a system."

---

20  Recently there emerged a number of works on the relationship between phenomenology and QBism. For more details on phenomenological teachings of the perspectival character of experience and connections to QBism, see, e.g., (Berghofer 2022, Chapter 15).
21  There are also other reasons why QBists should be interested in phenomenological investigations of experience. For instance, QBists argue that (quantum) measurement outcomes are the very experiences of the observing subject. However, QBists are physicists with no formal training in phenomenology or philosophy of mind. So when asked what precisely the experience looks like that is supposed to correspond, for instance, to the outcome of a spin-up/spin-down measurement or what exactly an instrument-mediated experience represents, answers remain vague. It is precisely here that phenomenologists could come to the rescue. Also, it has been speculated that quantum states in the QBist picture can be interpreted as features of what phenomenologists refer to as the intentional horizon of experience (de la Tremblaye 2020 and Pienaar forthcoming).

Although Rovelli does not offer a full-fledged reconstruction, these two postulates play a prominent role in the recent reconstructions offered in Höhn (2017) and Höhn & Wever (2017), where they are the rules 1 and 2, respectively (in a technically more precise formulation). What makes the two postulates particularly interesting for us is that they are simple, non-mathematical, and seem well-suited to be translated into experiential terms (see below). Furthermore, these principles seem to capture the somewhat paradoxical character of quantum mechanics. As Rovelli points out, "[t]here is an apparent tension between the two statements": "If there is a finite amount of information, how can we keep gathering novel information?" (Rovelli 2018, 7). Importantly, this tension is only apparent and resolved by the fact that previously acquired information can become irrelevant. As an illustration, Rovelli offers the example of a spin-1/2 particle that we first send through a z-oriented Stern-Gerlach apparatus and then through an x-oriented apparatus. First we gain information about $L_z$ (angular momentum in z-direction) and then we acquire information about $L_x$. However, when we gain the $L_x$ information, we lose the $L_z$ information.

Now, the question we are interested in is whether we can reformulate these postulates such that we avoid the factive term of "information." More precisely, given the QBist focus on the notion of experience, our question is whether we can reformulate in terms of experience. Our results from Subsection 4.1 suggest that we can reformulate straightforwardly, for instance as follows:

*Postulate 1\** (Limited perspective). There is a maximum amount of experiential input corresponding to a distinctive perspective from which an object can be experienced.

*Postulate 2\** (Unlimited perspectives). It is always possible to acquire a new perspective on an object.

Postulate 1* corresponds to the thesis discussed in the previous subsection that experiences can only provide a limited perspective. Postulate 2* corresponds to the thesis that these perspectives are complementary. Our main motivation for reformulating Rovelli's information-theoretic principles in terms of experience was that this non-factive language coheres better with *ψ-doxastic* interpretations. Relatedly, our principles are metaphysically more parsimonious and moderate. Informational reconstructions presuppose that there are external systems from which information can be extracted. Our principles, by contrast, are about the nature of experience. If they are true, they apply even if there is nothing external that can be extracted. I used the term "object" so that there is a straightforward analogy to Rovelli's "system." But object can be understood in the phenomenological sense of intentional object; a term that is metaphysically neutral, merely expressing that an experience is directed at something (whether this be an external mind-independent object or not). Accordingly, our approach would be consistent with the analysis of Grinbaum (2017), according to which quantum reconstructions are stepping stones toward theories that "do away with the idea of entities" (see also Adlam 2022). It also coheres nicely with Mittelstaedt's claim that the success story of modern physics can be constructed as a story of abandoning metaphysical hypotheses (Mittelstaedt 2011).

Of course this is only a very rough sketch of what reconstructions in experiential terms could look like. However, I hope to have argued convincingly that since the project of QBism is most accurately spelled out in terms of non-factive mental states such as belief and experience, it is only reasonable to complement QBism with a reconstruction that is similarly formulated in terms that avoid factive concepts such as knowledge and information. More generally, this applies to any *ψ-doxastic* interpretation that is motivated by QRP. In the next and final section, we discuss whether there are options "in between" *ψ-epistemic* and *ψ-doxastic* interpretations.

**5. Reconstructions as a road toward *ψ-doxastic* interpretations?**

We understood *ψ-ontic* interpretations in the informal sense of saying that the wave function represents the ontic state of a physical system. This aligns well but is not committed to wave function realism. While the currently most popular interpretations in philosophy are all versions of *ψ-ontic* interpretations, recently it has been argued that the successful information-based reconstructions of the quantum formalism that emerged in the last two decades speak in favor of non-ontic interpretations of the wave function. Almost universally, this is understood as supporting a *ψ-epistemic* interpretation. A *ψ-epistemic* interpretation is typically understood as saying that the wave function represents one's *knowledge* of the underlying ontic state. In this paper, I have emphasized that *ψ-epistemic* interpretations are not the only alternative to *ψ-ontic* interpretations. This is important for two reasons. First, one may find *ψ-epistemic* interpretations unattractive for various reasons addressed above, and it would be a false dichotomy to believe that these are the only two options. Second, QBism, an interpretation that was borne out of QRP, is best described as what I called a *ψ-doxastic* interpretation. In Section 3.4, I argued that QBists should be interested in reconstructing the quantum formalism in non-factive terms, and in Section 4.2, I offered a first glimpse of what could be the starting point of an experience-based reconstruction. Now, in this final section, let us assume that we are on the right track. If (and of course this is a big "if") reconstructions in experiential terms are possible and we should interpret the wave function neither in a *ψ-ontic* nor a *ψ-epistemic* manner, is a *ψ-doxastic* interpretation the only option left? From an epistemological perspective, there is at least one further obvious option.

In epistemology, it is common to distinguish between the fact *p*, one's knowledge that *p*, one's belief that *p*, and one's *justification* for believing that *p*. Traditionally, it has been assumed that epistemic justification is the link between mere belief and knowledge. Obviously, facts are factive. If *p* is a fact, *p* is true. Knowledge is also factive. If one knows that *p*, *p* is true. Beliefs are not factive. One may believe that *p*, but *p* is false. Although more controversial, justification is also *not* factive. Most epistemologists would agree that one can be justified in believing *p* although *p* is not true. Importantly, however, justification is *objective* in the following sense: Given a certain epistemic situation (a subject has a specific experiential input and background beliefs), the subject is justified in believing some proposition *p*, independent of whether the subject actually believes that *p*.[22] Furthermore, it would be true for *any* subject in this epistemic situation that they are justified in believing that *p*. We may also say: If in a given epistemic situation a subject is justified to believe that *p*, epistemically speaking, the subject *should* believe that *p*, whether or not she actually does.

We can now see how this connects to the previous section. If we succeeded in reconstructing the quantum formalism from principles about the nature of experience, we could say that the quantum formalism has the following function: given a certain experiential input, it tells the experiencing subject what she should expect to experience next. This partially aligns with the QBists' focus on experience. But it does not align so well with their focus on belief. Importantly, QBists insist that the wave function does *not* represent what the subject *should* believe. It only represents (degrees of) belief. In this respect, my approach aligns almost perfectly with Markus Müller's results presented in (Müller 2020).

---

22  Here we are concerned with propositional justification (as opposed to doxastic justification). Basically, this corresponds to the distinction between "*having justification to believe that p* versus *justifiedly believing that p*" (Silva & Oliveira forthcoming). The former characterizes propositional justification, denoting whether a subject has justification to believe some proposition (whether or not the subject actually believes the respective proposition); the latter characterizes doxastic justification, denoting whether a subject's actual belief is justified.

Here Müller develops a "first-person-first" approach to science. According to this framework, the objective of science is not "to describe the objective evolution of a unique external world" but to answer the question: "'**What will I see next?**'" (Müller 2020, 2; emphasis in the original). While QBism presupposes the existence of an external world[23], Müller's sole starting point is what he calls the "observer state": "a mathematical formalization of the information-theoretic state of the observer, including its current observations and its memory (conscious and unconscious)" (3). For our topic this means that "[w]e should not think of the wave function as the 'configuration of the world' in a naive sense, but rather as a catalogue of expectations about what an agent will see next" (3). According to Müller, "[o]ne of the clearest arguments for this broadly 'epistemic' view comes from the recent wave of reconstructions of QT, which proves that the full complex Hilbert space formalism of QT can be derived from a few natural information-theoretic principles" (28).

Such an experience-centered first-person-first approach coheres nicely with an experience-first epistemology as developed in (Berghofer 2022). In this light, a reconstruction in experience-based terms as suggested in the previous section could be viewed as paving a way from epistemology to quantum mechanics. This suggests an interpretation of quantum probabilities according to which they are not degrees of belief (as QBists would have it) but objective degrees of epistemic justification. In terms of Bayesian probability theory, quantum probabilities may not be understood as subjective probabilities along the lines of Bruno de Finetti but rather as objective Bayesian probabilities in the Coxian sense according to which probabilities represent *reasonable* expectations. This is to say that "the primary meaning of probability" is its function to provide a "measure of reasonable expectation" (Cox 1946, 2).[24] In this picture, quantum mechanics is a machinery into which we feed some experiential input (encoded in the wave function), and as an output, we get the answer to the question of what we *should* believe to experience next, based on this experiential input. This output comes in form of objective degrees of epistemic justification. I consider it an advantage over QBism that in this "degrees of epistemic justification interpretation" (DEJI), objectivity enters from the get-go.[25] Of course, it would go beyond the scope of this paper to discuss in detail the advantages and shortcomings of DEJI. At this point, I only want to emphasize that there is room "in between" epistemic and doxastic interpretations of the wave function.

---

23 "A QBist assumes the existence from the world from the outset" (Fuchs 2023, 92).
24 Interestingly, Cox also discusses "[t]he relation between expectation and experience" (Cox 1961, vii), pointing out that "since probability is relative to an experience that is never complete, it is always subject to change by new experience" (Cox 1946, 10).

Above we saw that Fuchs described the development of QBist thinking about quantum states as follows: "*Knowledge → Information → Belief → Pragmatic Commitment*" (Fuchs 2002). An option apparently not considered is that the wave function represents objective degrees of epistemic justification. Here is a more complete and fine-grained list of different options[26] for how to interpret the wave function:

>  *ψ-ontic*: the wave function represents the ontic state of the physical system.
> 
> *ψ-epistemic*: the wave function represents one's knowledge about the state of the physical system.
> 
> *ψ-justificational*: the wave function represents objective degrees of epistemic justification regarding the contents of one's future experiences.
> 
> *ψ-doxastic*: the wave function represents one's degrees of belief about the contents of one's future experiences.

---

25  DEJI shares many systematic similarities with Healey's pragmatist approach to quantum theory. For Healey, quantum theory "is a source of objectively good advice about *how* to describe the world and what to believe about it as so described. This advice is tailored to meet the needs of physically situated, and hence informationally-deprived, agents like us" (Healey 2022, Section 4.3). The main differences between DEJI and Healey's pragmatism are that in the latter there is no focus on the notion of experience and no talk in terms of epistemic justification. It is to be noted that when it comes to epistemic justification, I'm not a pragmatist but champion the following "objective" view: The degree of propositional justification a subject has for believing some proposition *p* is an objective matter of fact that is independent of the subject's goals, wishes, or desires.

26  A further option would be to combine these options. For instance, Hance, Rarity, and Ladyman (2022) have recently pointed out that "there is no reason to suppose that a one-one map between the wavefunction and the ontic state rules out that the wavefunction represents knowledge" (see also Luc 2023). This is to say that interpretations can be both *ψ-ontic* and *ψ-epistemic* (although not in the strict sense introduced in Harrigan and Spekkens 2010). One approach to quantum mechanics that exploits this fact is Steven French's phenomenological interpretation based on the work of Fritz London and Edmond Bauer (French 2023, Chapter 10).

The future will show whether reconstructions in terms of experience are possible and whether *ψ-justificational* interpretations can be established. This paper may serve as a stepping stone in this direction.

**Conclusion**

This paper focused on the relationship between the quantum reconstruction program and *ψ-doxastic* interpretations. Since QRP is in tension with *ψ-ontic* interpretations, and since *ψ-epistemic* interpretations are in tension with technical results and conceptual reflections, engaging QRP with *ψ-doxasticism* is a natural move. However, reconstructing the quantum formalism in factive terms such as knowledge and information does not fit well with *ψ-doxasticism*. Accordingly, I argued for a reconstruction in non-factive terms. Since QBism, the most prominent *ψ-doxastic* interpretation, centers around the concept of experience, I suggested a reconstruction in terms of experience. In Section 4.1, I discussed how phenomenology in Husserl's tradition approaches the notion of experience. In this light, in Section 4.2, I showed how Rovelli's information-theoretic principles could be reformulated in experiential terms. Of course, this constitutes only a first step toward investigating whether QRP should aim at reconstructions in non-factive experiential terms.

**Acknowledgments:** A previous version of this paper was discussed in a reading group session at the Center for Philosophy of Science at the University of Pittsburgh. I'm particularly grateful for the feedback from Adam Koberinski, Eleanor Knox, Lucy Mason, John Norton, and David Wallace. I'm most indebted to Philip Goyal for convincing me of the significance of the quantum reconstruction program and for many helpful discussions and clarifications. This research was funded in part, by the Austrian Science Fund (FWF) [10.55776/P36542].

**Statements and declarations**

The author declares that there are no conflicts of interest.